\documentclass[12pt]{iopart} %**HERE

\usepackage{graphicx}% Include figure files
\usepackage{bm}% bold math
\usepackage{iopams}  
\usepackage{epstopdf}
\begin{document}

\title[Plasmon scattering from holes]{Plasmon scattering from holes: from single hole scattering to Young's experiment}

\author{T Wang$^1$, E Boer-Duchemin$^1$,  G Comtet$^1$, E Le Moal$^1$, G Dujardin$^1$, A Drezet$^2$ and S Huant$^2$}
\address{$^1$ Institut des Sciences Mol\'eculaire d'Orsay (ISMO),
CNRS Univ Paris-Sud, 91405 Orsay, France\\ }
\address{$^2$Institut N\'eel, CNRS, Grenoble, France}
\ead{Elizabeth.Boer-Duchemin@u-psud.fr}%µ**HERE

\begin{abstract}
In this article, the scattering of surface plasmon polaritons (SPPs) into photons at holes is investigated.  A local, electrically excited source of SPPs using a scanning tunnelling microscope (STM) produces an outgoing circular plasmon wave on a thick (200~nm) gold film on glass containing holes of 250, 500 and 1000~nm diameter.  Fourier plane images of the photons from hole-scattered plasmons show that the larger the hole diameter, the more directional the scattered radiation.  These results are confirmed by a model where the hole is considered as a distribution of horizontal dipoles whose
relative amplitudes, directions, and phases depend linearly on the local SPP electric field.  An SPP-Young's experiment is also  performed, where the STM-excited SPP-wave is incident on a pair of 1~$\mu$m diameter holes in the thick gold film.  The visibility of the resulting fringes in the Fourier plane is analyzed to show that the polarization of the electric field is maintained when SPPs scatter into photons. From this SPP-Young's experiment, an upper bound of $\approx$~200~nm for the radius of this STM-excited source of surface plasmon polaritons is determined.
\end{abstract}

 %A CHANGER
\submitto{\NT}
                         
\maketitle %% required

\section{Introduction}
Surface plasmon polaritons (SPPs) are intensely studied for their use in potential nanophotonic applications as their  electromagnetic fields can be confined to dimensions much smaller than the wavelength of  light\cite{Barnes2003}.  SPPs are a key element in the extraordinary optical transmission (EOT) of light\cite{Ebbesen1998} through arrays of holes of subwavelength diameter in opaque metal films, a phenomenon which has generated much excitement and  fundamental and applied research\cite{Genet2007,Garcia-Vidal2010}.  Despite this intense activity, an understanding of the \emph{scattering} of SPPs into \emph{photons} at holes remains incomplete.

In order to understand the scattering of plasmons into photons at holes and better understand EOT, a well-controlled experiment is necessary.  In such an experiment, surface plasmon polaritons must be excited on the sample away from the hole, and the plasmons must have the opportunity to propagate to and interact with the hole. An experiment using this geometry has recently been reported\cite{Rotenberg2012} in which the authors focused on the scattering of an SPP plane wave from a single subwavelength hole into forward and radial plasmon waves. To our knowledge, however, an extensive study on the SPP scattering from a single subwavelength hole into photons has not been realized before our work.
Single holes have also been investigated by \emph{directly} exciting the hole  and measuring the transmitted light in the far field\cite{Prikulis2004,Alaverdyan2007,Yi2012},  in the near-field\cite{Yin2004,Chang2005,Rindzevicius2007} or using  a leakage radiation microscope\cite{Baudrion2008}.    A scanning near-field microscope (SNOM) tip in illumination mode has also been used to investigate single subwavelength holes\cite{Brun2003,Sonnichsen2000}.  However, the spatial and angular distribution of the light scattered from SPPs at single holes,   has not been studied until now.

Young's experiment---the observation of an interference pattern when an opaque screen perforated by two holes is placed before a light source---has been investigated under various different forms involving plasmons\cite{Sonnichsen2000,Zia2007,Derouard2007,Haefele2012,Schouten2005,Kuzmin2007, Ravets2009,Lalanne2005,Aigouy2007,Gan2007,Pacifici2008,Pacifici2008a,Guebrou2012,Guebrou2012a}. In particular, an ``all SPP'' version has been demonstrated where the ``holes'' are replaced by two metal stripe waveguides\cite{Zia2007}. Hole pairs have also been optically excited simultaneously\cite{Haefele2012,Schouten2005}, as well as individually\cite{Kuzmin2007, Ravets2009},   demonstrating  the existence of plasmon propagation between slits in such experiments\cite{Schouten2005,Kuzmin2007, Ravets2009,Lalanne2005}. The light scattered from the ends of a locally excited nanowire may also be considered a type of Young's experiment\cite{Kolesov2009,Zhang2013}. Again, however, the interference between the light scattered at two holes from propagating surface plasmons has never been previously examined.  Such an experiment is important as it provides a method for studying the coherence of SPPs.

%The scanning tunneling microscope (STM) is a powerful tool for the local, low energy, electrical excitation of both localized and propagating surface plasmons\cite{Wang2011,Bharadwaj2011,Wang2012}.  While the inelastic tunneling excitation of surface plasmons has been attributed to the statistical fluctuations of the tunnel current\cite{Schneider2010,Douillard2011} (ck), giving rise to a broadband spectrum of emitted light, this physical process is far from understood.  What is the physical origin of this broadband spectrum?  Are the created plasmons homogeneously or heterogeneously broadened?  Does the propagation length or this broadband spectrum best represent the coherence of STM-excited plasmons?

In this article, we investigate the scattering of surface plasmon polaritons into photons at single and double holes on a 200~nm-thick gold film. These SPPs are excited \emph{electrically} and \emph{locally} with a scanning tunnelling microscope (STM), producing an outgoing circular plasmon wave\cite{Wang2011,Bharadwaj2011}. This local excitation,  the ability to precisely position the excitation source and the absence of any background light from the excitation  are essential for these experiments. For the single hole experiment, diameters of 250, 500 and 1000~nm are considered. The scattered light at the holes is seen to be directional along the tip-hole axis and this directionality increases with hole diameter. For the double hole case, we see that the visibility of the resulting interference pattern varies as a result of excitation position due to the  polarization of the STM-excited plasmons.  Simulations where the hole scattering is considered as a series of in-plane coherent dipoles are in good agreement with the experimental results.  This work demonstrates a novel method for studying the coherence properties of surface plasmon polaritons and allows us to estimate an upper bound for the size of the excitation source.   
%\section{Experiment}

\section{Experimental methods}
The sample used consists of a 200~nm-thick (i.e., opaque) gold film deposited on glass.  Widely spaced single and pairs of holes with diameters of  250, 500 and 1000~nm are etched in the film by focused ion beam lithography.  The SPP-excitation on the gold film is carried out using an ambient STM coupled to an inverted optical microscope equipped with a x100 oil immersion objective (numerical aperture NA$=1.45$)\cite{Wang2011,LeMoal2013}.  Photons produced by the scattering of SPPs at the single holes and at the hole pairs are collected below the sample and focused onto a cooled charge-coupled device (CCD) camera.  An extra lens may be added in order to image the Fourier plane on the CCD camera. The collected light may also be analyzed with a spectrometer.  For all real space and Fourier space images shown in the following, the STM parameters are $I_{tunnel}$=~6~nA, $V_S$=~2.8~V and the integration time of the CCD camera is 60~s.  The STM tip used is made of electrochemically etched tungsten.

%\begin{figure}[htb]
%\includegraphics{single3}
%\caption{ a) Experimental set-up. A scanning tunneling microscope operating in air (Veeco Bioscope/Nanoscope IV) is combined with an inverted optical microscope (Zeiss Axiovert 200). The sample is a thin gold film (6, 20, 35, 50 or 70~nm) deposited on a $\sim$0.17~mm-thick glass coverslip (refractive index $n$=1.52). Light is detected through the sample using a cooled CCD camera (PIXIS 1024B, Princeton Instruments).  The extra  lens in front of the CCD camera is added for   Fourier plane imaging, otherwise the real space images are recorded.   Alternatively,  spectra may be obtained using a spectrometer  (Jobin-Yvon, Triax 190). b) ``Blunt'' tungsten tip made by alternating current electrochemical etching.  The radius of curvature is estimated to be $\sim$400~nm.  c) ``Sharp'' tungsten tip made by direct current electrochemical etching.  Radius of curvature: $\sim$70~nm.  c) Silver tip.  Radius of curvature: $\sim$400~nm. The red circles denote the estimated radii of curvature and the scale bars are 500~nm long.}
%\label{fig:exp_set-up}
%\end{figure}

\section{Single hole scattering}

\begin{figure}[htb]
\centering
 {\includegraphics[width=0.5\linewidth]{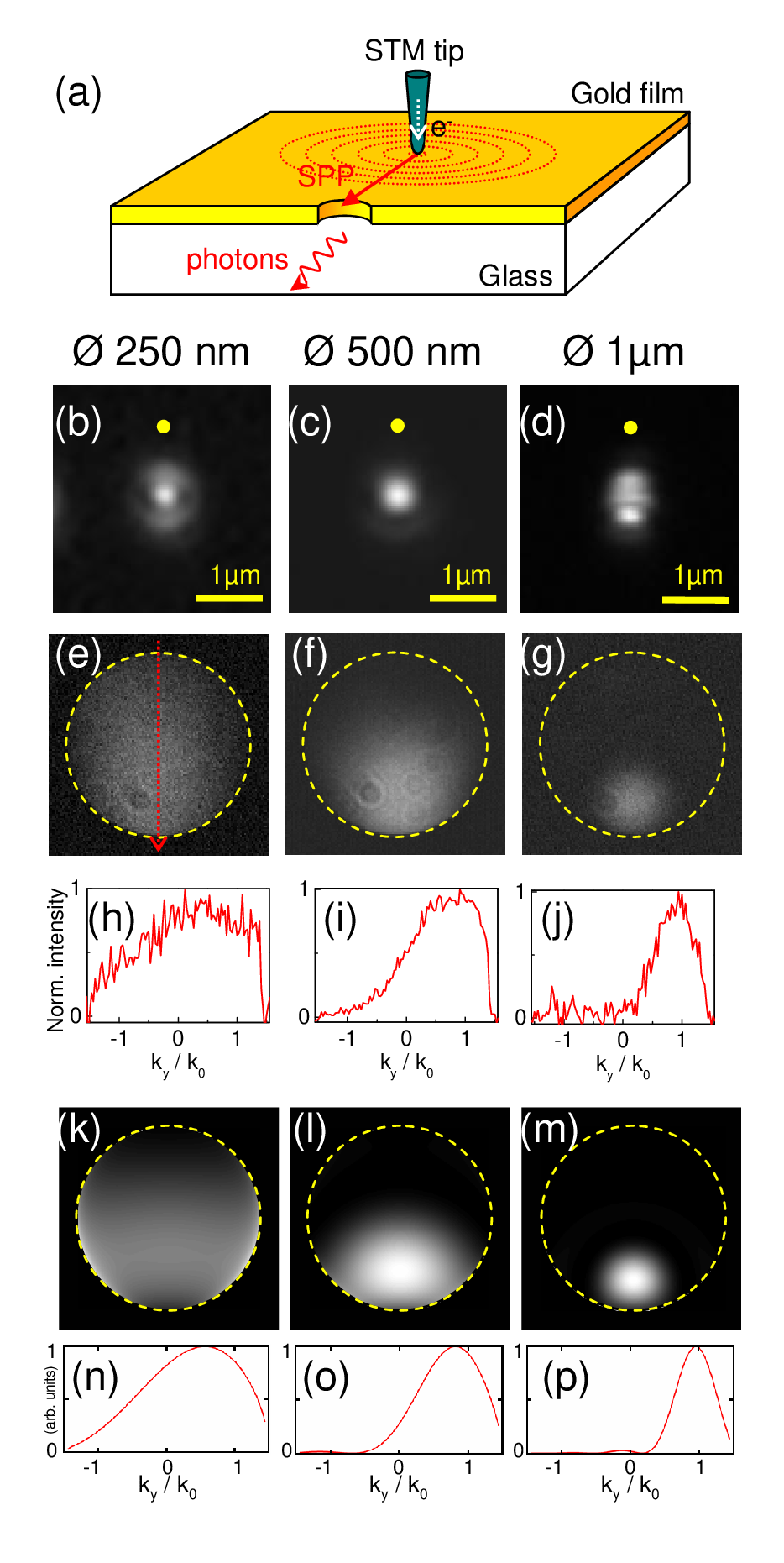}}
%{\includegraphics{Figure1a_1p}}
\caption{Single hole scattering: (a) Sketch of the experiment. The STM tip is positioned 1~$\mu$m from the hole (yellow dot in parts (b) to (d)) and excites an outgoing circular plasmon wave on the thick (200~$\mu$m) gold film (STM parameters $I_{tunnel}$=~6~nA and $V_S$=~2.8~V).  When the SPP wave interacts with the hole, the emitted photons are collected below the substrate. (b)-(d)  Real space images for hole diameters of 250, 500 and 1000~nm.  The real space image clearly varies as the hole diameter increases. (e)-(g) Fourier plane results and (h)-(j) corresponding cross-sections  obtained along the vertical axis of the figure (see the dotted red line in part (e)). $k_y$ is the $y$-component of the wave vector and $k_0$ the free space wave vector. As the hole diameter increases the directionality of the scattered light becomes more and more pronounced. Single hole simulation results: (k)-(m) Fourier plane images and (n)-(p) corresponding cross-sections  obtained along the vertical axis of the figure.  The hole diameters are (k)~250, (l)~500 and (m)~1000~nm.  As in the case of the experimental results, as the hole diameter increases the directionality of the scattered light becomes more and more pronounced.}
\label{fig:single}
\end{figure}
Figure~\ref{fig:single}(a) shows the principle of the single hole experiment.  The STM tip excites a circularly propagating plasmon wave (on the gold-air interface)  which upon reaching the hole is scattered into photons. 
%These photons are then collected through the transparent substrate.  
Figures~\ref{fig:single}(b)-(d) show the real plane images obtained during such an experiment for  three holes of different diameter.  In each case the STM tip excitation position is denoted by the yellow dot in the figure. 

The real space image varies as a function of hole diameter.  For the subwavelength 250~nm hole, the real space image consists of three bright spots aligned along the tip-hole axis, with the brightest spot centered on the hole.  This is reminiscent of a horizontal dipole above a glass substrate\cite{Sick2000}.  The result for the 500~nm hole consists of a single bright spot centered on the hole and the largest (1~$\mu$m) diameter again gives rise to a three spot pattern along the tip-hole axis.  This time the brightest spot is the one that is farthest from the tip excitation position.  Note that this result is reminiscent of the prolate shape observed in \cite{Sonnichsen2000} whose orientation depends on the polarization of the SNOM excitation light.

Figures~\ref{fig:single}(e)-(g) and the vertical cross-sections (h)-(j) show the corresponding Fourier plane images for the single hole experiments.  The differences due to hole diameter are even more remarkable in the Fourier plane.  A slight asymmetry is observed for the Fourier plane image of the 250~nm hole with a symmetry axis along the tip-hole direction.  This asymmetry becomes more and more pronounced as the size of the hole increases, with the radiation clearly forward-peaked near the air/glass critical angle for the largest sized hole (1~$\mu$m).

%Figure~\ref{fig:sim_single} shows the simulation results of Fourier plane images for the three different hole diameters.  A hole is modelled as a series of horizontal dipoles whose relative phases depend on the local phase of the incoming plane wave. 

Figure~\ref{fig:single}(k)-(q) shows the simulation results of Fourier plane images for
the three different hole diameters. A hole is modelled as a
distribution of horizontal dipoles $\vec{P}(\vec{r} )$ whose
relative amplitudes, directions, and phases at a point $\vec{r} =
[x,y]$ in the plane depend linearly on the local SPP in-plane  electric field
$\vec{E}(\vec{r})$ of the incoming SPP plane wave at the same
location. We have
$\vec{P}(\vec{r})=\alpha\vec{E}(\vec{r})$ where the
polarizability $\alpha$ is chosen constant for simplicity.  The
radiation field imaged in the Fourier plane of the high NA
objective is calculated by using the exact Green dyadic propagator
for the electromagnetic field in the non-paraxial regime\cite{Paulus2000,nanooptics,Tang2007,Hohenau2011},
and by summing over the dipole distribution in the hole. See the Appendix~A for further details.
As in the case of the experimental data, the emitted radiation becomes strongly peaked in the forward direction as the hole diameter increases.  This may be understood as a diffraction/interference phenomenon in which the emitted radiation interferes constructively in the forward direction and destructively otherwise.  Thus we may consider the scattering of plasmons from holes analogous to the scattering of light by particles, where the object's response to an optical excitation is considered multipolar, and retardation effects are taken into account.  It is these resulting phase differences which give rise to the directivity of the scattered light.

\section{Double hole scattering: Young's experiment}

\begin{figure}[htbp]
  \centering
 % {\includegraphics{double_col}}
 %{\includegraphics{double}}
 {\includegraphics[width=0.7\linewidth]{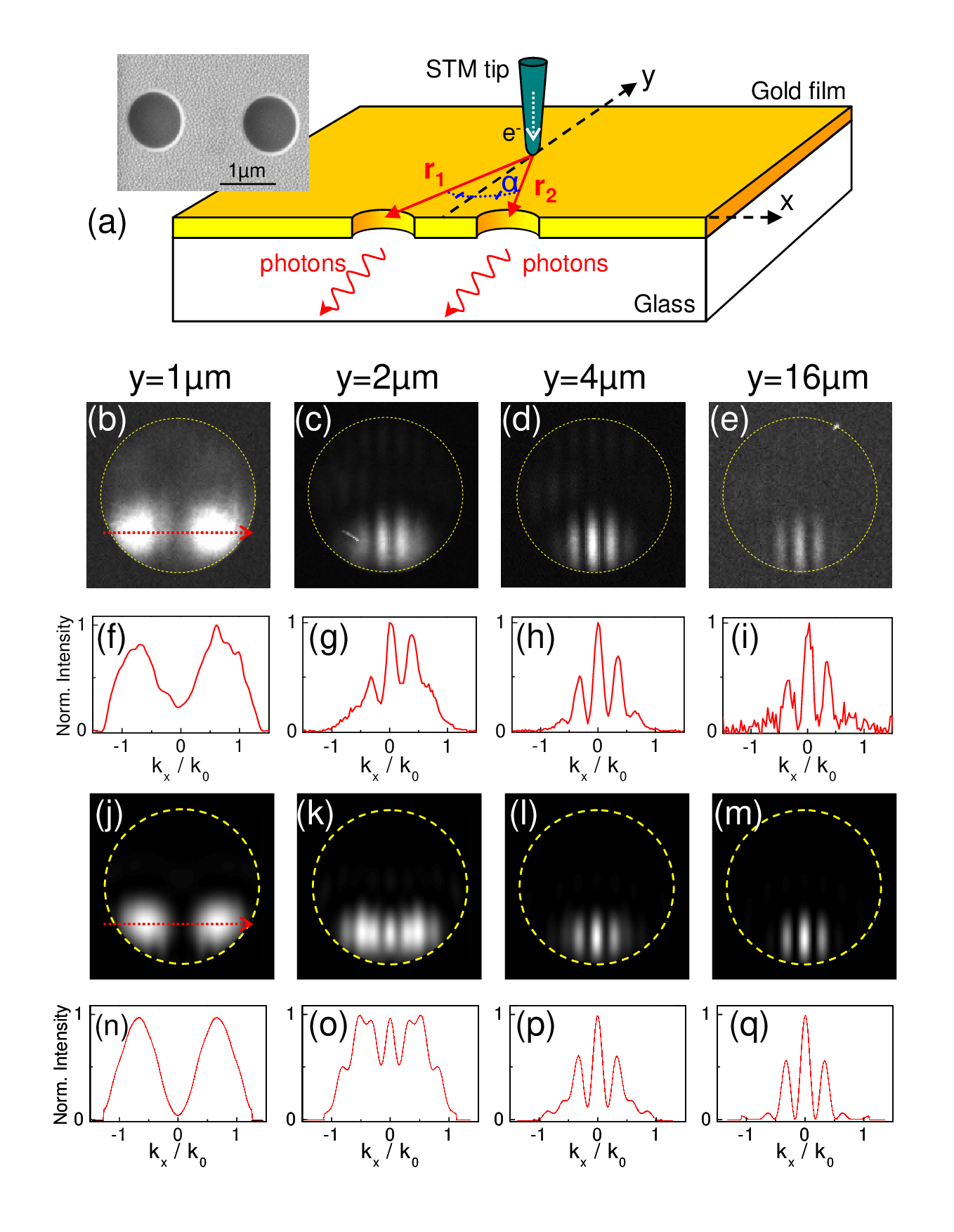}}
 %{\includegraphics{Figure2a_2m}}
  \caption{Hole pair scattering and interference---Young's experiment: (a) Principle of the experiment and scanning electron micrograph of the two holes; in the experiment the STM excites a circular plasmon wave on a thick Au film and the SPP wave scatters into photons at the two 1~$\mu$m diameter holes. (b)-(i) Fourier plane images and corresponding cross-sections obtained by collecting the emitted light below the substrate for tip excitation positions of (b), (f) 1~$\mu$m, (c), (g) 2~$\mu$m , (d),(h) 4~$\mu$m and (e), (i) 16~$\mu$m along the $y$-axis (see part (a)).  The cross-sections are obtained perpendicular to the fringes where the fringe intensity is maximal (see the dotted red line in part (b)). $k_x$ is the $x$-component of the wave vector and $k_0$ the free space wave vector. Note the decrease in the signal-to-noise ratio due to the fact that the excitation point is far from the holes in (i) ($y=16 \mu$m). (j)-(q)  Hole pair simulation results:  Two (plane) plasmon waves propagating from the ``tip'' located at (j), (n)~1~$\mu$m (k),(o)~2~$\mu$m, (l), (p)~4~$\mu$m and (m), (q)~16~$\mu$m  along the $y$-axis (see figure~\ref{fig:double}(a)) are incident on  two  1~$\mu$m diameter holes separated by 2~$\mu$m.  As in the experimental results, the calculated Fourier plane images show no fringes when the polarization of the two incoming plasmon waves is orthogonal (j),(n), and the visibility of the fringes increases as the ``tip'' is moved away from the holes and the polarization of the two incoming plasmon waves becomes more and more parallel. The agreement with experiment is best for larger values of $y$. This may be because effects such as the creation of plasmons  at one hole and their subsequent interaction  with the other hole would be more significant for smaller values of $y$ and are not included in the model.}
  \label{fig:double}
\end{figure}

In the next experiment a pair of 1~$\mu$m-diameter holes separated by 2~$\mu$m is used. In figure~\ref{fig:double}(a) the principle of the experiment is explained and a scanning electron microscope image of the hole pair is displayed.  The STM tip is positioned along the $y$-axis (i.e., the perpendicular bisector of the line joining the two holes) and a circular plasmon wave is excited on the Au film with the STM.  The SPP wavefronts travel a distance $|\vec{r}|$  to the holes before being scattered into the far field.  As in Young's double slit experiment, an interference fringe pattern will be observed in the Fourier plane if there is a fixed phase difference between the radiation from the two holes.  

From these experiments we  gain information on the plasmon source size, polarization and coherence.  Figure~\ref{fig:double}(b)-(i) shows Fourier plane images and the corresponding cross-sections of the resulting fringes when the STM tip is used to excite SPPs at different positions along the perpendicular bisector of the line joining the two holes.  As the tip is moved away from the two holes, a dramatic increase in the contrast or visibility is initially observed.  The visibility then stabilizes near a value of 1 for tip positions even further away on the $y$-axis. The visibility is defined as
\begin{equation}
V=\frac{I_{max}-I_{min}}{I_{max}+I_{min}}
\label{eqn:visibility1}
\end{equation}
 where $I_{max}$ and $I_{min}$ are the intensities corresponding the maximum and adjacent minimum of the fringes\cite{Hecht_book}.
 
The low value of the fringe visibility for tip positions close to the hole pair may be understood by considering the polarization of the excited plasmons.  As a first approximation we consider the two holes as point sources whose in-plane electric fields are in the direction of  SPP propagation (i.e., $\widehat{r}_{1}$ and $\widehat{r}_{2}$, see figure~\ref{fig:double}(a)).  Thus we have

\begin{equation}
\eqalign{\vec{E_{1}}=| \vec{E_{1}}|e^{i\phi_1}\, \widehat{r}_{1}\\
\vec{E_{2}}=| \vec{E_{2}}|e^{i\phi_2}\, \widehat{r}_{2}}
\end{equation}
where $| \vec{E_{i}}|$ and $\phi_i$ are the amplitude and phase respectively of the two point sources.  When these two sources interfere in the Fourier plane we get
\begin{equation}
%\fl \eqalign{I(k)&=(\vec{E_{1}}e^{-ik\frac{d}{2}}+\vec{E_{2}}e^{ik\frac{d}{2}})^{2}\cr
% &=| \vec{E_{1}}|^{2}+| \vec{E_{2}}|^{2}\\
% &+ \vec{E_{1}}\cdot\vec{E_{2}}^{*}e^{-ikd}+\vec{E_{1}}^{*}\cdot\vec{E_{2}}\,e^{ikd}}
\fl \eqalign{I(k)&=|\vec{E_{1}}e^{-ik\frac{d}{2}}+\vec{E_{2}}e^{ik\frac{d}{2}}|^{2}\cr
 &=| \vec{E_{1}}|^{2}+| \vec{E_{2}}|^{2}+ \vec{E_{1}}\cdot\vec{E_{2}}^{*}e^{-ikd}+\vec{E_{1}}^{*}\cdot\vec{E_{2}}\,e^{ikd}}
\end{equation}
where $d$ is the distance between the two holes and $k= \frac{2\pi }{\lambda_{0} } n\sin\theta$ is the coordinate in the Fourier plane (i.e., the in-plane component of the wave vector of the emitted radiation. $\lambda_{0}$ is the photon wavelength in free space and $n$ is the index of refraction and $\theta$ is the angle with respect to the optical axis). Thus, after averaging over a finite interval longer than the coherence time  and taking into account the correlations between the optical disturbances at each hole  we obtain
\begin{equation}
I(k)=I_{1}+I_{2}+ 2\cos(\alpha)\sqrt{I_{1}I_{2}}\left |\gamma_{12}(\tau)  \right |\cos(kd+\Delta \phi)
\label{eqn:I}
\end{equation}

with
%\begin{equation}
\begin{eqnarray}
I_{1}=|\vec{E_{1}}|^{2} \ \ \textrm{ and } \ \ \ I_{2}=| \vec{E_{2}}|^{2}
\\ [0.2cm]
\cos(\alpha)=\widehat{r}_{1}\cdot \widehat{r}_{2}=\frac{y^{2}-(d/2)^{2} }{y^{2}+(d/2)^{2} }
\end{eqnarray}
%\end{equation}
where  $\gamma_{12} (\tau)$ is the complex degree of coherence\cite{Born1999}, and is related to the ability of the light from the two holes to form interference fringes. $I_1$ and $I_2$ are the respective intensities at each hole. $\tau$ is a time interval equal to the path difference  between the source and the two holes divided by the velocity.  With the same plasmon wavefront arriving at the two holes at the same time (see figure~\ref{fig:double}(a)) we have $\tau=0$ and $\Delta \phi=0$.   The $\cos(\alpha)$ term is the result of the in-plane polarization of the source plasmons.

\begin{figure}[htbp]
  \centering
  {\includegraphics{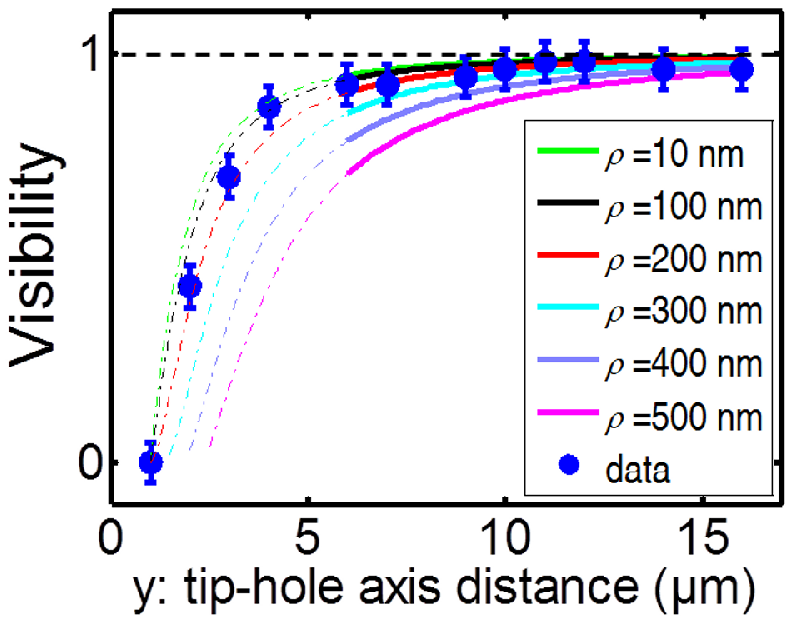}}
  \caption{Visibility as a function of tip-hole axis distance ($y$): the blue dots show the data obtained from figures such as figure~\ref{fig:double}(b)-(i) above.  The curves are obtained from equations~\ref{eqn:visibility2} and \ref{eqn:gamma} with $d=2$~$\mu$m and $\lambda_0$=700~nm; $\rho$ is the radius of the source. Note that equation~\ref{eqn:gamma} is only valid for $y>>d$, i.e., for tip-hole axis distances that are large as compared to the hole separation. }
  \label{fig:size}
\end{figure}

From the definition of the visibility (equation~\ref{eqn:visibility1}) and the above (equation~\ref{eqn:I}) and taking $I_1=I_2$ since the holes are equidistant from the source we obtain
\begin{equation} 
V (visibility)=\cos(\alpha)\ |\gamma_{12}(0)|=\frac{y^{2}-(d/2)^{2} }{y^{2}+(d/2)^{2} }|\gamma_{12}(0)|
\label{eqn:visibility2}
\end{equation}
Thus when the holes are 2~$\mu$m apart ($d/2$=1~$\mu$m) and  the excitation point is 1~$\mu$m away from the hole axis ($y=1~\mu$m) $\cos(\alpha)=0$ and the visibility falls to zero.  This is confirmed in figure~\ref{fig:double}(b) and (f) where no fringes are seen. It should be noted that while there is less and less overlap between the scattered light from the two holes as the tip is brought closer to them due to the directionality of the scattering, it is the polarization of the scattered light that causes the lack of interference fringes. On the other hand, when the tip is comparatively far from the holes as in figure~\ref{fig:double}(e) and (i) ($y=16~\mu$m), $\cos(\alpha)\cong1$ and the visibility is maximal.  This evolution of the visibility with the tip excitation position is shown in more detail in figure~\ref{fig:size} (blue data points).  These results clearly show that the light scattered at the two holes maintains the initial polarization of the incident plasmon wave.

Figure~\ref{fig:double}(j)-(m) shows the simulation results for the hole pair experiments.  Again the calculations agree well with the experimental data. The small discrepancies at large angle (i.e. large $k_x$, $k_y$)
 between the data and our model are possibly due to geometrical
 aberrations in the objective that are not taken into account in the
 simulation. Another source of error may be that no hole-reflected SPP wave is taken into account.
 
Not only do these experimental results tell us about the source polarization, but also about the source size.  If we approximate the source seen by the two holes as a disc and use the van Cittert-Zernike theorem\cite{Hecht_book, Born1999} we can determine the  degree of coherence $|\gamma_{12}(0)|$  at the two holes\cite{Thompson1957}.  This formalism is only valid when  both the source and hole separation are small compared to the tip-hole distance.  Once these conditions are satisfied, the degree of coherence  $|\gamma_{12}(0)|$ is equal to the absolute value of the normalized Fourier transform of the intensity function of the source, or more explicitly for a circular source :
\begin{equation} 
|\gamma_{12}(0)|=\frac{2J_1(\beta)}{\beta } \textrm{ with } \beta\approx\frac{2\pi}{\lambda_0}\frac{\rho d}{y}
\label{eqn:gamma}
\end{equation}

$\rho$ is the radius of the circular source, $d$ is the distance between the two holes, $\lambda_0$ is the wavelength and $y$ is the perpendicular distance from the source to the hole axis.  Of these variables, only the size of the source is unknown.  Thus from the data we can first determine  the visibility (via equation~\ref{eqn:visibility1}) then find the degree of coherence $|\gamma_{12}(0)|$ from equation~\ref{eqn:visibility2} and finally estimate an upper bound for the source size from equation~\ref{eqn:gamma}.  In figure~\ref{fig:size}, we have plotted the visibility for different values of the source radius $\rho$ and find a best fit to the data for the case where $\rho\approx200~$nm.  Note that all the curves for $\rho<200$~nm pass through the error bars of the data so that this is indeed only an upper bound for the effective source size.  The error introduced by the fact that the source is not strictly monochromatic may be shown to be on the order of 2\% (see Appendix~B). If the STM tip excitation position is no longer restricted to the $y$-axis (i.e., the perpendicular bisector of the hole axis), then the plasmon path difference for the source to each of the two holes is no longer zero and $\tau\neq0$.  In such an experiment, $|\gamma_{12}(\tau)|$ may be determined and the \emph{temporal} coherence of the STM-excited surface plasmons investigated.
%~\ref{sec:spectrum}).

\section{Conclusion}
In conclusion we have shown that the radiation from STM-excited SPPs scattered at holes becomes more and more directional as the hole size increases.  This effect has been reproduced using a dipolar model.  An SPP-Young's experiment has been performed and the visibility as a function of the excitation position investigated, demonstrating that the polarization is maintained when SPPs are scattered into photons at holes.  From this visibility data, an upper bound of $\approx$~200~nm on the SPP source size has been determined. Such a small, electrically-excited SPP source that generates no background excitation radiation is a unique tool for the study of SPP coherence, and quantum SPP properties such as wave particle duality and SPP coupling to quantum emitters\cite{Kolesov2009,Cuche2010, Mollet2012,Tame2013}.

\ack
We thank R. Marty, C. Girard and J.J. Greffet for fruitful discussions and J.F. Motte of the NANOFAB, Institut N\'eel, Grenoble for samples. This work is supported by the ANR project NAPHO (contract ANR-08-NANO-054) and the European STREP ARTIST (contract FP7 243421).
\appendix

\section*{\label{sec:model}Appendix A: Model}
\setcounter{section}{1}

%\section{Model}
In our model we consider a cylindrical hole of radius $a$ and height
$h$  in a metal film with permittivity
$\varepsilon_{\rm{metal}}$. The mathematical approach consists
of removing a cylinder of metal from the film of thickness $h$
and replacing it with an identical cylinder filled with air (i.e., with
permittivity
$\varepsilon_{\rm{air}}\simeq 1$).\\
The electric displacement field
$\mathbf{D}(M)$ at point $M$ in the region surrounding each hole is
given by a Lippman Schwinger integral
$\mathbf{D}(M)=\mathbf{D}_{\rm{SPP}}(M)+\int_{V}d^{3}x'\bar{\textbf{G}}_{\rm{film}}(M,M')(\varepsilon_{\rm{air}}-\varepsilon_{\rm{metal}})\mathbf{E}(M')$
where $\bar{\textbf{G}}_{\rm{film}}(M,M')$ is the total dyadic
Green tensor corresponding to the film without a hole, the integration
volume $V$ corresponds to the cylindrical region occupied by the
hole (filled with air) and $\mathbf{D}_{\rm{SPP}}(M)$ is the
incident SPP field propagating along the interface $z=0$ and existing
without the hole. In the transmitted region (i.e., in the substrate),
$\mathbf{D}_{\rm{SPP}}(M)\approx 0$ and only the volume integral
survives. Now, to a first-order (Born) approximation, we can write in
the transmitted region $\mathbf{D}(M)\simeq
\int_{V}d^{3}x'\bar{\textbf{G}}_{\rm{film}}(M,M')(\varepsilon_0-\varepsilon_1)\mathbf{E}_{\rm{SPP}}(M')$,
where $\mathbf{E}_{\rm{SPP}}$ is the incident unperturbed SPP
field. However, since the SPP field strongly decays in the metal
(penetration length $\simeq 10 nm$) the volume integral
evolves into a surface integral over the aperture area $S$:
$\mathbf{D}(M)\simeq\frac{i}{k_1}
\int_{S}d^{2}\mathbf{r}'\bar{\textbf{G}}_{\rm{film}}(M,\mathbf{r}',z'=0^-)(\varepsilon_0-\varepsilon_1)\mathbf{E}_{\rm{SPP}}(\mathbf{r}',z'=0^-)$,
where the coefficient $\frac{i}{k_1}$ arises from the integration of
the SPP exponential decay in the metal.  We point out that it is
mainly the in-plane field which contributes to the signal since in
the metal $|E_z|\ll |\mathbf{E}_{||}|$.\\
Finally, in the far-field the propagation through the microscope can
be taken into account by modifying the dyadic Green function. In the
Fourier plane of the objective the signal field is therefore to a
first approximation proportional to the structure factor defined by
\begin{eqnarray}
\mathbf{Q}[\mathbf{k}_{||}]=\int_{S}d^{2}\mathbf{r}'e^{-i\mathbf{k}_{||}\cdot\mathbf{r}'}\mathbf{E}_{\rm{SPP}}(\mathbf{r}').
\end{eqnarray}
which is calculated for the in-plane wave vector
$\mathbf{k}_{||}=2\pi/\lambda\cdot
n\sin{\theta}[\cos{\varphi}\hat{\textbf{x}}+\sin{\varphi}\hat{\textbf{y}}]$
($n$ is the oil index,  $\theta$ and $\varphi$ are the angle of
photon emission in the spherical coordinate system with symmetry
axis $z$). In our model we also take into account the modification
of this formula using the formalism developed by Tang et al.\cite{Tang2007}
for large numerical aperture microscope objectives.  The dependence of the results on wavelength is found to be weak in the wavelength range of interest.  Consequently, the calculations used the peak wavelength of the measured spectrum. 

\appendix
\section*{\label{sec:spectrum}Appendix B: Spectrum}
\begin{figure}[htbp]
  \centering
  %{\includegraphics{FigureS1bis(eps)}}
  {\includegraphics{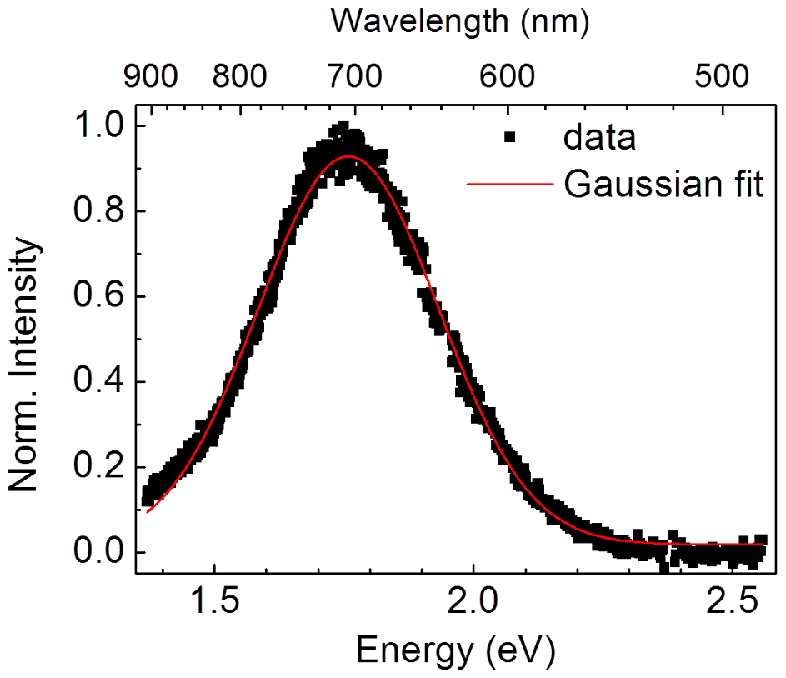}}
  \caption{Spectra of STM-excited plasmons scattered by a pair of 1~$\mu$m diameter holes in a 200~nm thick Au film. STM parameters are $I_{tunnel}$=~6~nA, $V_S$=~2.8~V and the integration time is 300~s.  The STM tip is located 2~$\mu$m from the hole axis, along its perpendicular bisector, i.e., $y=2$~$\mu$m. }
  \label{fig:spectra}
\end{figure}

\setcounter{section}{2}
%\section{Spectrum}
The error introduced by the fact that the source is not strictly monochromatic may be shown to be on the order of 2\%. The coherence length $L_c$ of a source may be determined from its spectral bandwidth  $\Delta\nu$ via the expression\cite{Goodman1985}
\begin{equation}
L_c=v_{plasmon}\sqrt{\frac{2\ln2}{\pi}}\frac{1}{\Delta\nu}.
\label{eqn:Lc}
\end{equation}
where $v_{plasmon}$ is the plasmon wave velocity. From a spectral measurement of the light scattered by a pair of 1~$\mu$m diameter holes (see figure~\ref{fig:spectra}) we obtain a value of $\approx$~2~$\mu$m for $L_c$.  Similarly, the degree of coherence as a function of path difference $l$ is given by the expression\cite{Goodman1985}
\begin{equation}
|\gamma(l)|=\exp(-\frac{\pi}{2}\frac{l^2}{L_c^2})
\label{eqn:gamma_gauss}
\end{equation}
Using the value of $L_c$ found from the measured spectrum and $l=\rho\approx200$~nm for the (maximum) path difference we obtain $|\gamma(200\rm{ nm})|\approx0.98$ (compared to a value of 1 that would be obtained with a strictly mono-chromatic source with infinite coherence length).  From equation~\ref{eqn:visibility2} in the main text we see that the resulting error on the visibility will be at most on the order of 2\%.

\section*{References}
%\newpage %Just because of unusual number of tables stacked at end
\bibliographystyle{unsrt}

\begin{thebibliography}{10}

\bibitem{Barnes2003}
W.~L. Barnes, A.~Dereux, and T.~W. Ebbesen.
\newblock Surface plasmon subwavelength optics.
\newblock {\em Nature}, 424(6950):824--830, 2003.

\bibitem{Ebbesen1998}
T.~W. Ebbesen, H.~J. Lezec, H.~F. Ghaemi, T.~Thio, and P.~A. Wolff.
\newblock Extraordinary optical transmission through sub-wavelength hole
  arrays.
\newblock {\em Nature}, 391(6668):667--669, February 1998.

\bibitem{Genet2007}
C.~Genet and T.~W. Ebbesen.
\newblock Light in tiny holes.
\newblock {\em Nature}, 445(7123):39--46, January 2007.

\bibitem{Garcia-Vidal2010}
F.~J. Garcia-Vidal, L.~Martin-Moreno, T.~W. Ebbesen, and L.~Kuipers.
\newblock Light passing through subwavelength apertures.
\newblock {\em Rev. Mod. Phys.}, 82(1):729--787, January 2010.

\bibitem{Rotenberg2012}
N.~Rotenberg, M.~Spasenovic, T.~L. Krijger, B.~le~Feber, F.~J. Garcia~de Abajo,
  and L.~Kuipers.
\newblock Plasmon scattering from single subwavelength holes.
\newblock {\em Phys. Rev. Lett.}, 108(12):127402, March 2012.

\bibitem{Prikulis2004}
J.~Prikulis, P.~Hanarp, L.~Olofsson, D.~Sutherland, and M.~Kall.
\newblock Optical spectroscopy of nanometric holes in thin gold films.
\newblock {\em Nano Lett.}, 4(6):1003--1007, June 2004.

\bibitem{Alaverdyan2007}
Y.~Alaverdyan, B.~Sepulveda, L.~Eurenius, E.~Olsson, and M.~Kall.
\newblock Optical antennas based on coupled nanoholes in thin metal films.
\newblock {\em Nat. Phys.}, 3(12):884--889, December 2007.

\bibitem{Yi2012}
J.~.~M. Yi, A.~Cuche, F.~de~Leon-Perez, A.~Degiron, E.~Laux, E.~Devaux,
  C.~Genet, J.~Alegret, L.~Martin-Moreno, and T.~W. Ebbesen.
\newblock Diffraction regimes of single holes.
\newblock {\em Phys. Rev. Lett.}, 109(2):023901, July 2012.

\bibitem{Yin2004}
L.~Yin, V.~K. Vlasko-Vlasov, A.~Rydh, J.~Pearson, U.~Welp, S.~H. Chang, S.~K.
  Gray, G.~C. Schatz, D.~B. Brown, and C.~W. Kimball.
\newblock Surface plasmons at single nanoholes in au films.
\newblock {\em Appl. Phys. Lett.}, 85(3):467--469, July 2004.

\bibitem{Chang2005}
S.~H. Chang, S.~K. Gray, and G.~C. Schatz.
\newblock Surface plasmon generation and light transmission by isolated
  nanoholes and arrays of nanoholes in thin metal films.
\newblock {\em Opt. Express}, 13(8):3150--3165, April 2005.

\bibitem{Rindzevicius2007}
Tomas Rindzevicius, Yury Alaverdyan, Borja Sepulveda, Tavakol Pakizeh, Mikael
  Kall, Rainer Hillenbrand, Javier Aizpurua, and F.~Javier Garcia~de Abajo.
\newblock Nanohole plasmons in optically thin gold films.
\newblock {\em J. of Phys. Chem. C}, 111(3):1207--1212, January 2007.

\bibitem{Baudrion2008}
Anne-Laure Baudrion, Fernando de~Leon-Perez, Oussama Mahboub, Andreas Hohenau,
  Harald Ditlbacher, Francisco~J. Garcia-Vidal, Jose Dintinger, Thomas~W.
  Ebbesen, Luis Martin-Moreno, and Joachim~R. Krenn.
\newblock Coupling efficiency of light to surface plasmon polariton for single
  subwavelength holes in a gold film.
\newblock {\em Opt. Express}, 16(5):3420--3429, March 2008.

\bibitem{Brun2003}
M.~Brun, A.~Drezet, H.~Mariette, N.~Chevalier, J.~C. Woehl, and S.~Huant.
\newblock Remote optical addressing of single nano-objects.
\newblock {\em Europhys. Lett.}, 64(5):634--640, December 2003.

\bibitem{Sonnichsen2000}
C.~Sonnichsen, A.~C. Duch, G.~Steininger, M.~Koch, G.~von Plessen, and
  J.~Feldmann.
\newblock Launching surface plasmons into nanoholes in metal films.
\newblock {\em Appl. Phys. Lett.}, 76(2):140--142, January 2000.

\bibitem{Zia2007}
Rashid Zia and Mark~L. Brongersma.
\newblock Surface plasmon polariton analogue to young's double-slit experiment.
\newblock {\em Nature Nanotech.}, 2(7):426--429, July 2007.

\bibitem{Derouard2007}
Marianne Derouard, Jerome Hazart, Gilles Lerondel, Renaud Bachelot,
  Pierre-Michel Adam, and Pascal Royer.
\newblock Polarization-sensitive printing of surface plasmon interferences.
\newblock {\em Optics Express}, 15(7):4238--4246, April 2007.

\bibitem{Haefele2012}
V.~Haefele, F.~de~Leon-Perez, A.~Hohenau, L.~Martin-Moreno, H.~Plank, J.~R.
  Krenn, and A.~Leitner.
\newblock Interference of surface plasmon polaritons excited at hole pairs in
  thin gold films.
\newblock {\em Appl. Phys. Lett.}, 101(20):201102, November 2012.

\bibitem{Schouten2005}
H.~F. Schouten, N.~Kuzmin, G.~Dubois, T.~D. Visser, G.~Gbur, P.~F.~A. Alkemade,
  H.~Blok, G.~W. Hooft, D.~Lenstra, and E.~R. Eliel.
\newblock Plasmon-assisted two-slit transmission: Young's experiment revisited.
\newblock {\em Phys. Rev. Lett.}, 94(5):053901, February 2005.

\bibitem{Kuzmin2007}
N.~Kuzmin, G.~W.~T. Hooft, E.~R. Eliel, G.~Gbur, H.~F. Schouten, and T.~D.
  Visser.
\newblock Enhancement of spatial coherence by surface plasmons.
\newblock {\em Opt. Lett.}, 32(5):445--447, March 2007.

\bibitem{Ravets2009}
S.~Ravets, J.~C. Rodier, B.~Ea Kim, J.~P. Hugonin, L.~Jacubowiez, and
  P.~Lalanne.
\newblock Surface plasmons in the young slit doublet experiment.
\newblock {\em Journal of the Optical Society of America B-optical Physics},
  26(12):B28--B33, December 2009.

\bibitem{Lalanne2005}
P.~Lalanne, J.~P. Hugonin, and J.~C. Rodier.
\newblock Theory of surface plasmon generation at nanoslit apertures.
\newblock {\em Phys. Rev. Lett.}, 95(26):263902, December 2005.

\bibitem{Aigouy2007}
L.~Aigouy, P.~Lalanne, J.~P. Hugonin, G.~Julie, V.~Mathet, and M.~Mortier.
\newblock Near-field analysis of surface waves launched at nanoslit apertures.
\newblock {\em Phys. Rev. Lett.}, 98(15):153902, April 2007.

\bibitem{Gan2007}
Choon~How Gan, Greg Gbur, and Taco~D. Visser.
\newblock Surface plasmons modulate the spatial coherence of light in young's
  interference experiment.
\newblock {\em Phys. Rev. Lett.}, 98(4):043908, January 2007.

\bibitem{Pacifici2008}
Domenico Pacifici, Henri~J. Lezec, Luke~A. Sweatlock, Robert~J. Walters, and
  Harry~A. Atwater.
\newblock Universal optical transmission features in periodic and quasiperiodic
  hole arrays.
\newblock {\em Opt. Express}, 16(12):9222--9238, June 2008.

\bibitem{Pacifici2008a}
D.~Pacifici, H.~J. Lezec, Harry~A. Atwater, and J.~Weiner.
\newblock Quantitative determination of optical transmission through
  subwavelength slit arrays in ag films: Role of surface wave interference and
  local coupling between adjacent slits.
\newblock {\em Phys. Rev. B}, 77(11):115411, March 2008.

\bibitem{Guebrou2012}
S.~Aberra~Guebrou, J.~Laverdant, C.~Symonds, S.~Vignoli, and J.~Bellessa.
\newblock Spatial coherence properties of surface plasmon investigated by
  young's slit experiment.
\newblock {\em Opt. Lett.}, 37(11):2139--2141, June 2012.

\bibitem{Guebrou2012a}
S.~Aberra~Guebrou, J.~Laverdant, C.~Symonds, S.~Vignoli, F.~Bessueille, and
  J.~Bellessa.
\newblock Influence of surface plasmon propagation on leakage radiation
  microscopy imaging.
\newblock {\em Appl. Phys. Lett.}, 101(12):123106, September 2012.

\bibitem{Kolesov2009}
Roman Kolesov, Bernhard Grotz, Gopalakrishnan Balasubramanian, Rainer~J.
  Stoehr, Aurelien A.~L. Nicolet, Philip~R. Hemmer, Fedor Jelezko, and Joerg
  Wrachtrup.
\newblock Wave-particle duality of single surface plasmon polaritons.
\newblock {\em Nat. Phys.}, 5(7):470--474, July 2009.

\bibitem{Zhang2013}
Yang Zhang, Elizabeth Boer-Duchemin, Tao Wang, Benoit Rogez, Genevieve Comtet,
  Eric Le~Moal, Gerald Dujardin, Andreas Hohenau, Christian Gruber, and
  Joachim~R. Krenn.
\newblock Edge scattering of surface plasmons excited by scanning tunneling
  microscopy.
\newblock {\em Opt. Express}, 21(12):13938--48, June 2013.

\bibitem{Wang2011}
T.~Wang, E.~Boer-Duchemin, Y.~Zhang, G.~Comtet, and G.~Dujardin.
\newblock Excitation of propagating surface plasmons with a scanning tunnelling
  microscope.
\newblock {\em Nanotechnology}, 22(17):175201, April 2011.

\bibitem{Bharadwaj2011}
Palash Bharadwaj, Alexandre Bouhelier, and Lukas Novotny.
\newblock Electrical excitation of surface plasmons.
\newblock {\em Physical Review Letters}, 106(22):226802, June 2011.

\bibitem{LeMoal2013}
Eric Le~Moal, Sylvie Marguet, Benoit Rogez, Samik Mukherjee, Philippe
  Dos~Santos, Elizabeth Boer-Duchemin, Genevieve Comtet, and Gerald Dujardin.
\newblock An electrically excited nanoscale light source with active angular
  control of the emitted light.
\newblock {\em Nano Letters}, 13(9):4198--205, September 2013.

\bibitem{Sick2000}
B.~Sick, B.~Hecht, and L.~Novotny.
\newblock Orientational imaging of single molecules by annular illumination.
\newblock {\em Phys. Rev. Lett.}, 85(21):4482--4485, November 2000.

\bibitem{Paulus2000}
M.~Paulus, P.~Cay-Balmaz, and O.~J.~F. Martin.
\newblock Accurate and efficient computation of the green's tensor for
  stratified media.
\newblock {\em Phys. Rev. E}, 62(4):5797--5807, October 2000.

\bibitem{nanooptics}
Lukas Novotny and Bert Hecht.
\newblock {\em Principles of nano-optics}.
\newblock Cambridge University Press, 2006.

\bibitem{Tang2007}
Wai~Teng Tang, Euiheon Chung, Yang-Hyo Kim, Peter T.~C. So, and Colin J.~R.
  Sheppard.
\newblock Investigation of the point spread function of surface plasmon-coupled
  emission microscopy.
\newblock {\em Opt. Express}, 15(8):4634--4646, April 2007.

\bibitem{Hohenau2011}
A.~Hohenau, J.~R. Krenn, A.~Drezet, O.~Mollet, S.~Huant, C.~Genet, B.~Stein,
  and T.~W. Ebbesen.
\newblock Surface plasmon leakage radiation microscopy at the diffraction
  limit.
\newblock {\em Opt. Express}, 19:25749--25762, 2011.

\bibitem{Hecht_book}
Eugene Hecht.
\newblock {\em Optics}.
\newblock Addison-Wesley, 1990.

\bibitem{Born1999}
Max Born and Emil Wolf.
\newblock {\em Principles of Optics}.
\newblock Cambridge University Press, 7th edition edition, 1999.

\bibitem{Thompson1957}
B.~J. Thompson and E.~Wolf.
\newblock Two-beam interference with partially coherent light.
\newblock {\em J. Opt. Soc. Am.}, 47:895, 1957.

\bibitem{Cuche2010}
A.~Cuche, O.~Mollet, A.~Drezet, and S.~Huant.
\newblock "deterministic" quantum plasmonics.
\newblock {\em Nano Letters}, 10(11):4566--4570, 2010.

\bibitem{Mollet2012}
Oriane Mollet, Serge Huant, Geraldine Dantelle, Thierry Gacoin, and Aurelien
  Drezet.
\newblock Quantum plasmonics: Second-order coherence of surface plasmons
  launched by quantum emitters into a metallic film.
\newblock {\em Phys. Rev. B}, 86(4):045401, July 2012.

\bibitem{Tame2013}
M.~S. Tame, K.~R. McEnery, S.~K. Oezdemir, J.~Lee, S.~A. Maier, and M.~S. Kim.
\newblock Quantum plasmonics.
\newblock {\em Nat. Phys.}, 9(6):329--340, June 2013.

\bibitem{Goodman1985}
J.~W. Goodman.
\newblock {\em Statistical Optics}.
\newblock Wiley, 1985.

\end{thebibliography}

\end{document}